\documentclass[twocolumn,showpacs,preprintnumbers,amsmath,amssymb]{revtex4}
\usepackage{graphicx}
\usepackage{graphicx}
\usepackage{dcolumn}
\usepackage{bm}
\begin{document}
\title{A hybrid model for fusion at deep sub-barrier energies}
\author{Ajit Kumar Mohanty}
\email{ajitkm@barc.gov.in}
\affiliation{Nuclear Physics Division, Bhabha Atomic Research Centre, 
Mumbai 400 085, India}
\date{\today}

\begin{abstract}
{
A hybrid model where the tunneling probability is estimated
based on both sudden and adiabatic approaches has been 
proposed to understand the heavy ion fusion phenomena at deep sub-barrier energies. 
It is shown that under certain approximations, it amounts to tunneling through two barriers:
one while overcoming the normal Coulomb barrier (which is of sudden nature) along the radial
direction until the repulsive core is reached and thereafter through an adiabatic barrier along
the neck degree of freedom while making  transition from a 
di-nuclear to a mono-nuclear regime through shape relaxation.
A general feature of this hybrid model is 
a steep fall-off of the fusion cross section, sharp increase
of logarithmic derivative L(E) with decreasing energy  
and the astrophysical S-factor
showing a maxima at deep sub-barrier energies particularly for near symmetric systems.
The model can explain the experimental fusion measurements for several systems
ranging from near symmetric systems like $^{58}Ni+^{64}Ni,~ 
^{58}Ni+^{58}Ni$ and $ ^{58}Ni+^{69}Y$
to asymmetric one like $^{16}O+^{208}Pb$  
where the experimental findings are very surprising. Since the second tunneling is along the neck
co-ordinate, it is further conjectured that deep sub-barrier fusion supression may not be
observed for the fusion of highly asymmetric projectile target combinations where adiabatic
transition occurs automatically without any hindrance. The recent deep sub-barrier fusion cross section
measurements of $^{6}Li+^{198}Pt$ system supports this conjecture.}
\end{abstract}
\maketitle

\section{INTRODUCTION}

Fusion cross sections of two  heavy nuclei at sub-barrier energy
have been studied extensively for last several years, since
it was realized that the experimental measurements are enhanced by several orders
of magnitude over the predictions of a simple barrier penetration model (BPM)
at energies near and below the Coulomb barrier \cite{bal,das1}. In this BPM picture, 
fusion reaction at sub-barrier energies is governed
by the tunneling through the Columb barrier followed by an absorption inside
the barrier which is often simulated through an incoming wave boundary condition.
Since BPM model is one dimensional in nature, it fails to explain sub-barrier fusion
enhancement and broad spin distribution for several systems as  heavy ion fusion is
a complex process involving
tunneling in multi-dimensions. 
Although several theoretical models (both macroscopic and microscopic)  have been proposed
to account for this large sub-barrier fusion enhancement, the most  
successful model that has emerged out of these studies is the description of fusion
within a coupled channel framework where the presence of couplings to various low lying   
inelastic and transfer channels are treated explicitly. Under certain approximation,
the channel couplings result in a distribution of barriers, one or more of which
having height less than the original Coulomb barrier, thus giving large enhancement.   
Therefore, a better understanding of the fusion
process followed through the distribution of barriers which can be obtained from the
second derivative of the product of the fusion cross section with energy \cite{rowley}.
In most experiments, fusion cross sections have been measured down to the $mb$ level
and the coupled channel calculations have been quite successful in reproducing the
general trends of the measured yields. 
While efforts are still being put
to improve upon the techniques of these calculations to understand the 
nature of fusion enhancement  
and the associated barrier
distribution, some of the early $Ni$ induced measurements 
at extreme sub-barrier energies \cite{jiang1,jiang2} 
(cross section below the $\mu b$ level for systems like 
$^{64}Ni+^{64}Ni$ and  $^{60}Ni+^{89}Y$)
have brought out new surprises which show supression
in fusion cross section with respect to the same coupled channel calculations that
explains the enhancement at sub-barrier energies.
Subsequent measurements of fusion excitation functions for many other systems
like $^{64}Ni+^{100}Mo$, $^{28}Si+^{64}Ni$, $^{48}Ca+^{96}Zr$, $^{16}O+^{208}Pb$,
and $^{28}Si+^{30}Si$ \cite{jiang3,jiang4,trotta, das2,jiang5}, together with analysis
\cite{jiang6,jiang7,jiang8,jiang9,tak1} suggest that fusion hindrance is a general
phenomena expected to occur for all the systems at extreme sub-barrier energies
irrespective of the $Q$ value of the reactions. 
The most striking feature of these measurements is sharp increase of  the logarithmic 
derivative $L(E)=d[ln (\sigma E)]/dE$
with decreasing energies which can not be reproduced with normal
coupled channel calculations \cite{jiang6,hagino,sastry}. 
These calculations also fail to explain the behavior of $S$-factor which shows
a maxima at deep sub-barrier energies particularly for (near) symmetric systems.

To resolve these anomalies, several models have been proposed. One of the earlier
model which has been
successful in explaining the fusion suppression for most of the systems was first
proposed by Misicu and Esbensen \cite{mis1,mis2} where coupled channel calculations
are carried out using a modified ion-ion potential that includes a repulsive core
in addition to the normal double-folding potential with M3Y interaction. The
resulting potential is much more shallower as compared to standard double-folding 
potential. Both shallower pocket and thicker potential barrier result in fusion
hindrance at deep sub-barrier energies. Based on this potential, coupled channel calculations
have also been carried out for symmetric light system like $^{16}O+^{16}O$ to extrapolate $S$ factor
at deep sub-barrier energies which is
 of interest in astrophysical studies \cite{esben1}
and also to explain fusion hindrance of highly asymmetric $^{16}O+^{208}Pb$ system
\cite{esben2}. 
However, the above model provides only part of the explanation, but can not be
the complete picture as the shallow model has no potential pocket to trap higher
angular momenta and also fusion is not possible below the cut-off energy which is a technical
drawback. For example, in case of
$^{64}Ni+^{64}Ni$ system, the minimum value
of the potential pocket $V_{m} \sim 85$ MeV which is about $9$ MeV below
the Coulomb barrier. This would imply that fusion cross section $\sigma_f$ will
vanish for $E_{cm}\le V_{m}$ although compound nucleus formation is still possible
as long as $(E_{cm}+Q)$ is positive i.e. compound nucleus can in principle
be formed down to a threshold energy of $E_{cm} \sim 49$ MeV.
The experimental measurements also do not suggest a sharp cut-off in $\sigma_f$
for $E_{cm} \le V_{m}$.
The approach of Misicu and Esbensen is based on the sudden picture
where the nuclear reaction takes place so rapidly that the colliding nuclei overlap
with each other without changing their density. Fusion of two nuclei has also been studied
in the adiabatic picture where fusion trajectory is obtained after optimizing in other collective
variables like neck and mass asymmetric degrees of freedom. Both adiabatic and sudden approaches
may lead to the similar results in the region where the colliding nuclei do not overlap
significantly.  In Ref. \cite{tak2}, the authors consider the fission like adiabatic 
potential energy surface with neck configuration after the colliding nuclei come in contact
with each other. Their model consists of a capture probability in the two body potential
pocket followed by the penetration of the adiabatic one body potential to reach a 
compound state after the touching configuration. Similarly, there are 
models based on density constrained time dependent Hartree-Fock formalism which are used to estimate
fusion cross sections  at deep sub-barrier
energies \cite{umar} with varying degrees of success. 

It is also found that the deep sub-barrier suppression patterns are different for different
systems. The precision measurement of fusion cross sections
for $^{16}O+^{208}Pb$ system at deep sub-barrier energies show a steep but almost
saturated logarithmic slope \cite{das2} unlike earlier $Ni$ induced reactions where
logarithmic slope increases with decreasing energy. This has been interpreted as the
quantum decoherence effect of the channel wave functions caused by the coupling to the
thermal baths \cite{das2,torres}. The adiabatic method of Ref. \cite{tak2} has not been 
applied to $^{16}O+^{208}Pb$ system  and also the shallow potential model of Ref. \cite{mis1}
can not explain this suppression pattern unless coupling to transfer channel is included with
increased coupling strength \cite{esben2}. Recently, an improved coupled channel method
has been suggested \cite{tak3} to explain the fusion supression pattern
 of $Ni+Ni$ and $^{16}O+^{208}Pb$ systems. The use of the variable damping factor simulates
a smooth transition between the two-body and the adiabatic one-body state. While the deep 
sub-barrier suppression seems to be a general feature for many systems, another recent
measurements of the fusion cross
section for $^{6}Li+^{198}Pt$ system \cite{ara} shows no suppression although data exists for both
above and deep sub-barrier energies. It is interesting to note that 
a normal coupled channel calculation using breakup coupling with standard ion-ion 
potential can explain the above experimental observations quite well.
 Therefore, the fusion reaction study at
sub-barrier energy has now become a controversial topic and requires further investigations.

In this work,  we propose a hybrid coupled channel  model based on both sudden and adiabatic
pictures as follows. Initially, the colliding partners are required to overcome 
the conventional Coulomb barrier through tunneling along the radial direction.
After tunneling this sudden
barrier  at $r=R_b$, the di-nuclear system rolls down upto the bottom
of the pocket at $r=R_m$. Since beyond $r<R_m$, the sudden potential becomes repulsive, the 
radial motion becomes slower and the di-nuclear system undergoes shape relaxation making
a transition to the mono-nuclear regime that subsequently leads to complete fusion. 
It will be shown in this work
that the shape relaxation involves a second tunneling through an adiabatic barrier along the 
neck degree of freedom. The coupled channel calculations based on this double
penetration model is able to explain the deep sub-barrier suppression pattern
for several systems including $^{16}O+^{208}Pb$ experimental data. Although the present
formalism is quite similar in spirit to that of two step model of Ref.\cite{tak2}, 
the second tunneling in this work is calculated explicitly along the neck direction. It is argued
that for highly asymmetric projectile target combinations, the tunneling along the neck degree
is not essential and the transition to adiabatic trajectory occurs automatically once the
system overcomes the sudden Coulomb barrier. This feature, at least qualitatively, may explain
why $^{6}Li+^{198}Pt$ system does not show any deep sub-barrier fusion hindrance.

The paper is organized as follows. In section II, we take two examples
of ion-ion potentials, one based on Akuyz-Winther (AW) parameterization and second one, a recently
introduced M3Y+Repulsion (M3YR) double folding potential \cite{mis1,mis2}.
These potential models are 
sudden in nature. The adiabatic aspect is incorporated by adding an extra term due to
neck formation which is estimated using a simple macroscopic model of nuclear shape evolution.
It is shown that the adiabatic potential has a barrier
along the neck direction which the fusing system needs to overcome while making a transition
from a di-nuclear to a mono-nuclear regime. 
The fusion cross sections are estimated
using both adiabatic and sudden potential in one dimension. 
In section III, we propose a coupled
channel formalism based on a double penetration model which explains various aspects of 
experimental measurements at deep sub-barrier energy.
Finally, conclusions are presented in section IV. 

\section{ION-ION POTENTIAL}
The ion-ion potential is one of the important factor that governs the reaction
mechanism at sub-barrier energy. Attempts have been made to learn about the nuclear part
of the ion-ion potential at short distances from the measured fusion cross sections at deep 
sub-barrier energies using inversion technique \cite{hagino1}. Although the inverted potentials
for many systems are found to be thicker than phenomenological potentials and may partly explain
fusion hindrance, the inversion procedure is based on the assumption
that the experimental data can be fully explained
on the basis of coupled channel calculations and ignores any additional 
dynamical effects which may be playing important role at short distances. 
We will discuss about one such missing dynamical component in the subsequent section.
In the following, we consider only the ion-ion potentials which are of sudden and adiabatic
nature. The purpose here is to examine how much the deep sub-barrier fusion
process is affected by the shape of the ion-ion potential in the nuclear interior region.

\subsection{Sudden potential} 

As part of the sudden ion-ion potential, we use both Akuyz-Winther (AW) and
the M3Y+Repulsion double-folding potential (M3YR). The parameterized form of
AW potential is given by \cite{esben2}
\begin{eqnarray}
V_N(r)=\frac{-16 \gamma~\bar R~a}{1+exp\left \{\left(r-R_1-R_2-\Delta R\right)/a
         \right \}},
\end{eqnarray}
where $\Delta R$ is an adjustable parameter used to reproduce the barrier
height of double-folded potential with normal M3Y interaction. Here 
$\gamma=0.95$ MeV/fm$^2$ is the nuclear surface tension co-efficient , 
$R_i=1.2A_i^{1/3}-0.09$ fm, the diffuseness
parameter $a=0.63$ fm,  and $\bar R = R_1R_2/(R_1+R_2)$. 

In M3YR model, the standard double-folding potential is calculated using the integral
\begin{eqnarray}
V_N(r)=\int d \bf r_1 \int d \bf r_2 \rho_1(\bf r_1) \rho_2(\bf r_2)v(\bf r_{12})
\end{eqnarray}
where $r_{12}=\bf r_1+\bf r_2$ and $v(\bf r_{12})$ represents the M3Y effective
nucleon-nucleon interaction. The authors in Ref.\cite{mis1,mis2} simulated
a repulsive core by using an effective  contact interaction
\begin{eqnarray}
v_{rep}(\bf r_{12})=V_{rep}\delta(\bf r_{12}).
\end{eqnarray} 
The procedure followed is same as that of given in Ref. \cite{uegaki} where 
the double-folding integral is calculated using same radius parameter, but
with a sharp density profile, characterized by a smaller diffuseness parameter $a_{rep}$.
The strength of $V_{rep}$ of the repulsive interaction is obtained from the condition that
the nuclear potential at the origin $V_N(r=0)=(A_a~K)/9$ where $A_a$ is the mass number of the 
smaller nucleus and $K$ is the nuclear incompressibility factor.

\begin{figure}
\begin{center}
\includegraphics[scale=.4]{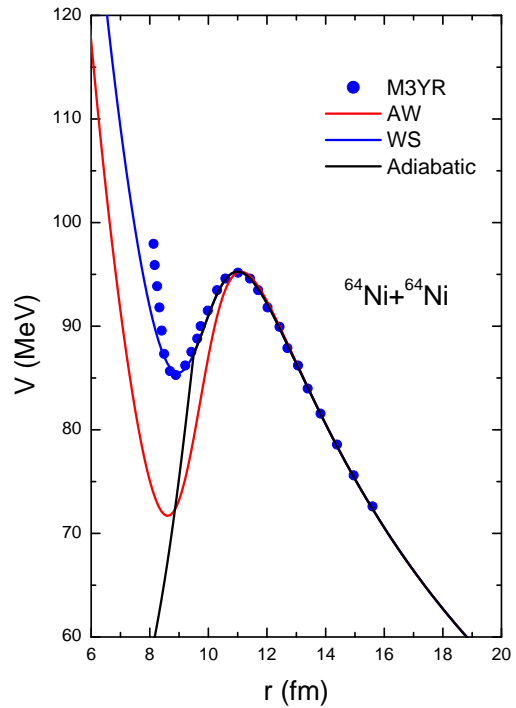}
\caption{ The total potential $V$ versus inter-nuclear distance $r$
for a typical $^{64}Ni+^{64}Ni$ system. The filled circles are calculations based
on M3YR potential taken from Ref. \cite{mis1}. The red curve 
represents the AW parameterization with $\Delta R=0.23$ fm. The blue curve is the
Wood-Saxon parameterization that reproduces the M3YR potential in the
region of interest. The WS parameters are $V_0=53$ MeV, $a_0=0.68$ fm, and $r_0=1.22$ fm respectively. 
The black curve represents the adiabatic potential ($R_a=9.5$, see text for detail).}
\label{fig1}
\end{center}
\end{figure}

Fig. \ref{fig1} shows the total potential (sum of nuclear and Coulomb) as the function of inter nuclear
distance $r$ for a typical $^{64}Ni+^{64}Ni$ system. The red curve represents the
potential based on the AW parameterization. The filled circles are calculation based on M3YR
potential (data points are taken from Ref. \cite{mis1}). The blue curve is the Wood-Saxon (WS) parameterization
that reproduces the M3YR potential in the region of interest. For M3YR potential,
we use this WS parameterization in calculations for convenience. As can be seen from 
Fig. \ref{fig1}, the parameters are so adjusted that both AW and M3YR potentials result in same barrier height
, but differ significantly in the nuclear interior region. The M3YR potential is more shallower as well as
more thicker as compared to the AW potential. These two aspects of M3YR potential result in 
the reduction of fusion cross section at deep sub-barrier energies. As mentioned before,
these potentials are of sudden nature where the shape of the nucleon density distributions are
kept frozen. This approximation is valid if the relative motion
is fast enough so that there is no time for internal rearrangements. When the relative motion slows down,
the potential becomes adiabatic and the presence of
other degrees of freedoms (like neck and  mass asymmetry) can not be ignored. 
In the following, we consider an adiabatic correction to these sudden
potentials which is shown to be important at nuclear interior region.

\begin{figure}
\begin{center}
\includegraphics[scale=.4]{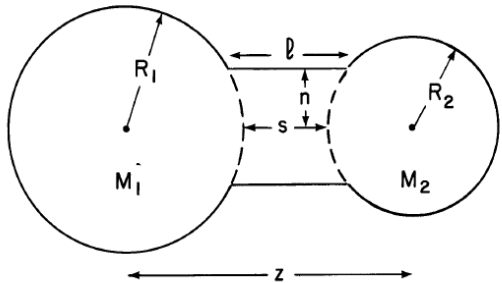}
\caption{A macroscopic representation of two nuclei connected by a cylindrical neck of length $l$ and radius
$n$. 
$R_1$, $R_2$ are the half density radii of two nuclei having mass
$M_1$ and $M_2$ and surface to surface distance $s$.
 Note that the centre to centre  distance $z$ and the radial distance $r$ have the
same meaning.}
\label{fig3}
\end{center}
\end{figure}

\subsection{Adiabatic potential}
We incorporate an adiabatic correction to the above sudden potential which becomes
important for $r<R_b$ where $R_b$ is the radius of the Coulomb barrier.
Fig. \ref{fig3} shows a macroscopic model where two nuclei are connected by a cylindrical neck.
This type of model has been used in Refs. \cite{swiat1,swiat2,ram} to study the fission and fusion dynamics
in two dimensional potential landscape comprising of elongation and neck degrees of freedom.
As shown in Fig. \ref{fig3}, $R_1$ and $R_2$ are the half density radii of the nuclei having masses 
 $M_1$ and $M_2$
, $s$ ($s=r-R_1-R_2$) being the surface to surface distance along the $z$ direction and $n$
 is the radius of the
cylindrical neck. The length of the cylinder $l$ can be written as \cite{swiat1},
\begin{eqnarray}
l \approx s+\frac{n^2}{2\bar R}~~~~;~~~~~~~~\bar R=\frac{R_1R_2}{R_1+R_2}
\end{eqnarray} 
The surface area (of interest) of the cylinder of length $l$ and radius $n$ that contributes to the
extra surface energy is given by (surface energy due to neck formation),

\begin{eqnarray}
V_{neck}(r,n)=2\pi\gamma (n l -n^2)
=2\pi\gamma(n s-n^2+\frac{n^3}{2\bar R}),
\label{v1}
\end{eqnarray} 
where $\gamma$ is the surface tension co-efficient in MeVfm$^{-2}$.

In the macroscopic model, elongation, mass asymmetry and neck degree of freedoms are considered
three independent collective variables which affect the fusion-fission dynamics. However, some macroscopic
models as in Refs. \cite{dav} and \cite{ada}
 consider the neck degree of freedom as a collective variable only for nuclear
interior region $r<R_b$ and completely omit the neck dynamics at the approach stage assuming that neck
is not an independent collective variable for $r > R_b$. Since the position
of the Coulomb barrier $R_b$ is much larger than the half density radius $R_1+R_2$, the overlap density in the
region at $r \sim R_b$ 
is very small as compare to the central saturation density. Such a di-nuclear system which is
formed purely due to geometrical overlap of the diffused nuclear surfaces at the approach stage
($r \ge R_b$) will only affect the nuclear potential, but can not grow into a fully relaxed di-nuclear
configuration. The role of neck dynamics has been investigated at the approach phase of reaction in Ref. \cite{ada}
using realistic mass parameters and friction coefficients. The analysis suggests that a macroscopic model
with neck variable leads to a relative motion of the nuclei that is similar to the motion in a potential
obtained in the frozen density approximation. 
Therefore, we do not expect any further correction in the
nuclear potential at the exterior region due to neck formation as the standard double folding potential
or its equivalent parameterization already accounts for the overlap correction  under the frozen density
approximation. 
However, as the colliding nuclei overcome the Coulomb barrier,
the nucleon density in the neck region approaches the saturation value resulting in a nuclear 
interaction which is repulsive 
at some point $R_m$ ($R_m \le R_1+R_2$). Since the relative motion in $r$ direction becomes slower,
the shape relaxation through neck growth (beyond the geometrical overlap $n_g$)
becomes meaningful at  $r \sim R_m$.
Therefore, for interior region ($r < R_b$), we introduce  
a correction term as given by Eq.\ref{v1}.

Finally, the adiabatic potential can be written as

\begin{eqnarray}
V_{ad}(r,\bar n)=V_N(r)+V_C(r)+V_{neck}(r,\bar n)
\label{ad} 
\end{eqnarray} 

where $V_{neck}(r,\bar n)$ is evaluated at an optimal neck radius $\bar n$ corresponding to a
fully relaxed shape configuration at the distance $r$. In otherwords, at each separation $r$,
the system undergoes shape relaxation
and makes a transition 
from a di-nuclear state (DNS) to a mono-nuclear state (MNS). As will be shown below, the transition
from the DNS to MNS is not automatic, the system finds a resistive path along the neck direction.

Using the dimensionless variables $\rho=s/(2\bar R)$ and  $\nu=n/(2\bar R)$, 
Eq.\ref{v1} can be written as,
\begin{eqnarray}
V_{neck}(\rho,\nu)=8\pi\gamma \bar R^2(\rho \nu -\nu^2+\nu^3)
\label{vneck}
\end{eqnarray} 

\begin{figure}
\begin{center}
\includegraphics[scale=.4]{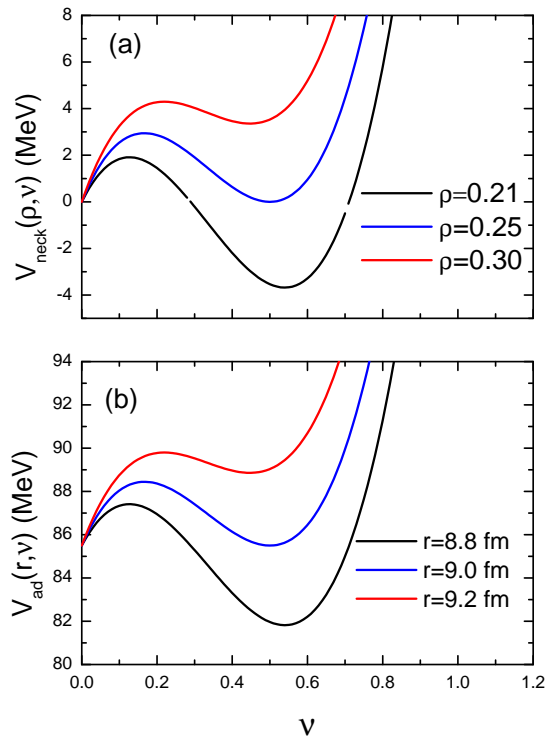}
\caption{ (a) The neck potential $V_{neck}(\rho,\nu)$ as a function of
$\nu$ for different values of $\rho$ for $^{64}Ni+^{64}Ni$ system.
The black curve is for $\rho=0.21$, the blue curve is for $\rho=0.25$ and 
the red curve
is for $\rho=0.30$ respectively. (b) The $V_{ad}(r,\nu)$
 as a function of $\nu$ corresponding to
three different $r$ values. The $R_a$ value is fixed at $9$ fm corresponding to the
distance where M3YR potential has the pocket.}
\label{fig4}
\end{center}
\end{figure}

The above expression is a simple cubic order polynomial which vanishes at $\nu=0$ and has a
 minimum at  
\begin{eqnarray}
\bar \nu=\frac{1+\sqrt{1-3\rho}}{3}.
\label{nubar}
\end{eqnarray}
as long as $\rho \le 1/3$. Between $\nu=0$ and $\bar \nu$, it is separated by a maxima at
\begin{eqnarray}
\nu_b=\frac{1-\sqrt{1-3\rho}}{3}.
\label{nub}
\end{eqnarray}

 Fig. \ref{fig4}(a) shows the plot of $V_{neck}(\rho,\nu)$
as a function of neck parameter $\nu$ at three  $\rho$ values for  a typical
$^{64}Ni+^{64}Ni$ system.  The system can undergo shape
relaxation from $\nu=0$ to $\nu=\bar \nu$ by overcoming the barrier at $\nu=\nu_b$. As can be seen from 
Fig. \ref{fig4},
for $1/3 \le \rho <  1/4$,
the height of the minimum  at $\nu=\bar \nu$ is higher
than the value at $\nu=0$. Since the state at $\nu=\bar \nu$
 is metastable, neck relaxation is not favored until 
the system reaches $\rho=1/4$ which corresponds to a inter nuclear separation of
 $r=R_1+R_2+\bar R/2$ (say $R_a$). At $\rho=1/4$ (i.e. $r=R_a$), 
$V_{neck}$ at $\nu=0$ and $V_{neck}$ at $\nu=\bar \nu$ are degenerate (both are zero). So, neck relaxation is possible i.e.
the system can make transition from the di-nuclear state at $\nu=0$ to a mono-nuclear
 state at $\nu=\bar \nu$ after overcoming
the barrier at $\nu=\nu_b$. 
In fact, for $\rho < 1/4$, $V_{neck}$ at $\nu=\bar \nu$
is always lower than $V_{neck}$ at $\nu=0$ so that neck relaxation is possible for $r\le R_a$.
Therefore, we can define $R_a$ as the adiabatic distance at which the
di-nuclear state in principle can make transition to a mono-nuclear state through shape
relaxation. The value of $R_a$ varies depending on how $R_i$ is defined i.e.  whether
it represents  half density or hard sphere
radius. However, we would like to treat $R_a$ as variable by redefining the surface to
surface distance  $s=r-(R_a-\bar R/2)$ so that $s=\bar R/2$
 at $r=R_a$ or $\rho=1/4$. This is an alternate and
convenient definition  which
ensures that the neck relaxation always begins at $r=R_a$. 

Like Fig. \ref{fig4}(a), Fig. \ref{fig4}(b) shows the plot of total
adiabatic potential $V_{ad}=V_N(r)+V_C(r)+V_{neck}(r,\nu)$ as a function of $\nu$ at 
three $\rho$ or $r$  values. We consider M3YR potential for nuclear part and fix 
$R_a$ at $9.0$ fm. The basic difference between Fig \ref{fig4}(a) and Fig. \ref{fig4}(b) is only the
shift in base values due to $V_N$ and $V_C$ contribution, where as there is no change in the barrier
height along the neck direction. 
If the collision energy is such that the system is able to overcome the neck barrier, the di-nuclear state will
make a transition to the mono-nuclear state and will attain
an optimal neck configuration. The black curve in Fig. \ref{fig1} shows the plot
of adiabatic potential as a function of $r$.     
Since neck potential is metastable for $r>R_a$, 
we set $V_{neck}(r,\bar \nu)=0$ for $r>R_a$ and estimate $V_{neck}(r,\bar \nu)$ 
for $r\le R_a$ which adds a negative correction to Eq. \ref{ad}. For the adiabatic
plot in Fig. \ref{fig1}, we have taken M3YR for $V_N$
and fixed $R_a=9.5$ fm. Note that the potential in 
Fig.\ref{fig4}b represents the adiabatic potential $V_{ad}(r,\nu)$
in the neck direction where as the black curve in fig.\ref{fig1} 
represents the adiabatic potential as a function of
$r$ for an optimum neck opening ($\nu=\bar \nu)$ (see Eq. \ref{v1}).  
It may be mentioned here that the choice of M3YR for $V_N$ part is purely arbitrary. The adiabatic correction
can be added to the AW parameterization as well. It is only to show
that since the correction begins at $r\le R_a$, the adiabatic corrected potential is no longer
repulsive and also the potential pocket does not exist any more. The fusion reaction is still possible
by tunneling through this adiabatic potential even for $E < V_m$ which was problematic with pure M3YR potential.
Another important aspect is that the adiabatic potential is always thiner than its sudden counterpart.

\subsection{Fusion in one dimension}
To demonstrate how the shape of the nuclear potential in the interior region affects the deep
sub-barrier fusion cross section, we first consider a simple one dimensional barrier penetration model (BPM), although
coupled channel calculations are carried out later on to explain the experimental data. It may
be mentioned here that even in the coupled channel calculations, the deep sub-barrier fusion cross section
is sensitive only to the lowest eigen barrier and the cross section follows mostly the calculations of an one
dimensional penetration model corresponding to the lowest eigen barrier.
The fusion in BPM can be written as 

\begin{equation}
\sigma_f=\sum_l \sigma_l=\frac{\pi}{k^2}\sum_l (2l+1) T_l(E),
\label{fus}
\end{equation}
where $k$ is the relative wave number and $T_l(E)$
is the tunneling probability which can be estimated using the WKB approximation,
\begin{eqnarray}
T_l(E)=\frac{1}{ 1+ exp(2S_l) },
\label{wkb}
\end{eqnarray}
where $S_l$ is the classical action given by,
\begin{eqnarray}
S_l = \frac{2\mu}{\hbar^2} \int_{r_1}^{r_2} \sqrt{V_l(r)-E } dr,
\end{eqnarray}
and 
\begin{eqnarray}
V_l(r)= V_C(r)+V_N(r)+\frac{l(l+1) \hbar^2}{2\mu r^2}.
\end{eqnarray} 
Under the parabolic approximation, Eq.(\ref{wkb}) can also be estimated using Hill-Wheeler
expression \cite{hill},
\begin{eqnarray}
T_l(E)=\left [ 1+ exp\left (\frac{2\pi}{\hbar \omega} (V_b^l-E) \right ) \right ]^{-1},
\label{hill}
\end{eqnarray}
where 
\begin{eqnarray}
V_b^l= V_b+\frac{l(l+1) \hbar^2}{2\mu R_b^2},
\label{bar1}
\end{eqnarray} 
and $V_b$ being  the $s$-wave barrier.
Knowing $\sigma_f$, we can estimate two sensitive quantities, the logarithmic derivative $L(E)$ and the
astrophysical $S$ factor which are given by,

\begin{equation}
L(E)=\frac{d[ln(E\sigma)]}{dE}~~~~;~~~~
S(E)=E\sigma e^{2 \pi (\eta-\eta_0)},
\label{astro}
\end{equation}
where $\eta=Z_1Z_2e^2/(\hbar v)$ is the Sommerfeld parameter , $v$ is the beam velocity and $\eta_0$ is a normalization
factor.

\begin{figure}
\begin{center}
\includegraphics[scale=.4]{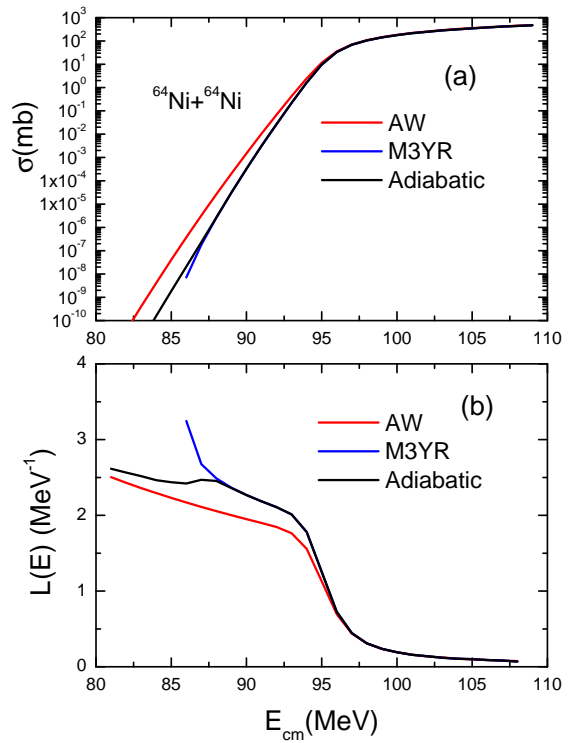}
\caption{(a) Fusion cross section as a function of $E_{cm}$ 
for the same $^{64}Ni+^{64}Ni$ system calculated using one
dimensional barrier penetration model. The red curve is based on the AW
parameterization, the blue curve is due to the WS parameterization of the
M3YR potential. The black curve is due to the adiabatic
potential as discussed in the text.
 (b) The corresponding $L(E)$ factors as a function of $E_{cm}$ for
the above system.}
\label{fig2}
\end{center}
\end{figure}

We estimate tunneling probability using WKB approximation (Eq. \ref{wkb}) 
for sub-barrier energy and Eq. \ref{hill} for energy above the Coulomb
barrier. Fig. \ref{fig2} shows the plot of fusion cross section and logarithmic 
derivative $L(E)$ for $^{64}Ni+^{64}Ni$ system as a function of energy. Being
one dimensional in nature, we do not expect this model to explain actual experimental data.
It is shown here only to demonstrate the basic difference between AW type of potential which is 
commonly used in many coupled channel calculations and the M3YR double folding potentail
with short distance repulsive correction which has been proposed recently in Ref. \cite{mis1} 
to explain deep sub-barrier suppression. As expected, the M3YR potential being thicker
and shallower shows fusion suppression as compared to AW potential (see blue and red
curves in fig. \ref{fig2}a). In case of AW potential, the $L(E)$ asymptotically saturates
with decreasing energy where as calculations with M3YR potential shows a sharp increase
in $L(E)$ with decreasing energy,  (see corresponding red and blue curves in Fig. \ref{fig2}b)
a similar trend as found in experimental measurements. However, this could be an artifact of shallow
nature of the M3YR potential which results in an abrupt cut-off of fusion cross section for energy
$E < V_m \sim 85$ MeV. In reality, the di-nuclear configuration has to move towards a 
mono-nuclear configuration followed by compound nucleus formation. Therefore, it will be
more appropriate to use an adiabatic potential which is of sudden nature for $R>R_a$ and truly
becomes adiabatic for $R \le R_a$ (see black curve in 
Fig. \ref{fig1}).  The black curve in Fig. \ref{fig2} shows the results obtained using the
adiabatic potential.
Note that using the adiabatic potential, it is now possible
to estimate fusion cross section for $E<V_m$ and also $L(E)$ shows saturation as one finds
in case of AW potential. In the present work, we have added adiabatic correction
to M3YR sudden potential only to show how the cut-off effect can be eliminated, although
adiabatic correction is also applicable to AW potential for $r<R_a$.

\section{FUSION HINDRANCE AT DEEP SUB-BARRIER ENERGY} 

In the conventional picture, fusion occurs automatically once the fusing system 
overcomes the Coulomb barrier.
While passage over this Coulomb barrier at $r=R_b$ is the necessary condition for fusion, 
it is not sufficient enough
to ensure that fusion will lead to compound nucleus formation. Further nuclear shape evolution is
necessary for the di-nuclear system to make transition from the initial sudden trajectory 
to final
adiabatic trajectory. If the system follows only the sudden trajectory, fusion leading to compound
nucleus formation is not possible as the potential becomes repulsive beyond $r < R_m$. 
Therefore,
it will be more appropriate to study fusion using the adiabatic trajectory (see Fig. \ref{fig1}) which has
been obtained minimizing over the neck coordinate, i.e. by allowing the system to relax from
$\nu=0$ to $\nu=\bar \nu$.
However, tunneling through the adiabatic trajectory alone may overestimate the fusion probability
particularly at deep sub-barrier energy
unless the barrier along the neck direction is taken into account. As shown in Fig.\ref{fig4}, 
a barrier along the neck direction exists between $(R_a-\bar R/2) \le r  \le R_a$ ($0 \le \rho \le 1/4$)
having a maximum height at $r=R_a$ and is given by (estimated  from Eq.\ref{vneck} at $\rho=1/4$ and
$\nu_b=1/6$),
\begin{eqnarray}
V_{neck}(R_a,\nu_b)=\frac{4\pi\gamma \bar R^2}{27}\approx 0.46 \gamma \bar R^2.
\label{vneckb}
\end{eqnarray}
and vanishes at $r=R_a-\bar R/2$.
Recall that
we have now three different distance parameters, $R_b$ (position of the Coulomb barrier), $R_a$
(the distance at which neck relaxation possible) and $R_m$ (the distance at which the sudden potential
has a minimum). Since, for fusion of stable nuclei, short distance repulsion and neck relaxation requires
significant nuclear density overlap, we expect $R_m \sim
R_a < R_b$.

Ideally, fusion process should be studied using the two dimensional 
adiabatic potential as defined in Eq.\ref{ad} which,  for a given mass asymmetry,
depends on both elongation and neck degree of freedoms. In the absence of such a dynamical tunneling
calculation in multi-dimensions, we propose here a simple two step model as follows. 
First the system
needs to tunnel through the Coulomb barrier at $R=R_b$ which can be estimated using 
WKB or Hill-Wheeler approximation.  Since for $r>R_a$ the potential is sudden in nature, 
this tunneling is same as what is being used under frozen density approximation and the di-nuclear
system follows the sudden trajectory until adiabatic transition becomes favorable at $r=R_a \sim R_m$.
Therefore, after overcoming the Coulomb barrier , the system rolls down upto the point $r=R_m$
beyond which the potential becomes repulsive. Since the fusing di-nuclear system can not
proceed in reducing $r$ direction beyond $R_m$, it undergoes shape relaxation at $r \sim R_m$
and is captured into the adiabatic valley that ultimately leads to compound nucleus formation. However,
for adiabatic transition (moving from $\nu=0$ to $\nu= \bar \nu$ along
neck direction), the system needs to overcome a barrier of height $V_{neck}(R_m, \nu_b)$  
at $R_m$ or at $R_a$ with available energy $(E-V_m)$
where $V_m$ is the pocket minimum. 

The fusion cross section can be written as,

\begin{equation}
\sigma_f=\frac{\pi}{k^2}\sum_l (2l+1) T_l^a(\epsilon)~T_l^s(E),
\label{fus_ad}
\end{equation}
where $T_l^s(E)$ is the tunneling probability through the sudden barrier at $R_b$ with energy
$E$ and $T_l^a(\epsilon)$ is the adiabatic tunneling probability in the neck direction at $r=R_m$ with energy
$\epsilon=(E-V_m)$. The above  factorization is approximate and 
can be justified for the fact that at the approach stage ($r\sim R_b$),
the relative motion is fast enough over the neck relaxation time and at the nuclear interior stage ($r \sim R_m$),
the neck grows rapidly as relative motion slows down. So, the total tunneling probability is 
written as the product of two factors,
the first tunneling along the relative degrees of freedom at $r=R_b$ 
and second tunneling along the neck degree of freedom at $r=R_a$ or $R_m$. For sub-barrier energy,
the second tunneling is almost unity, but provides hindrance for fusion at deep sub-barrier energies. 

For simplicity, we will use Hill-Wheeler approximation given by
Eq.\ref{hill} to estimate the above two tunneling probabilities. 
Under this approximation, it is now equivalent to say that the second tunneling needs to be calculated through
a  neck barrier ,
\begin{eqnarray}
V_n^l= V_m+V_{neck}(R_m,\nu_b)+\frac{l(l+1) \hbar^2}{2 I(R_m,\nu_b)},
\label{bar2}
\end{eqnarray} 
at energy $E$.
In the above equation, $V_m$ is the potential minimum at $R_m$ , $V_{neck}(R_m,\nu_b)$ is the maximum neck
barrier at $\nu=\nu_b$ and the last term represents the rotational energy of the mononuclear configuration
at $R=R_m$ and $\nu=\nu_b$. However, for simplicity, 
we use the same term $I=\mu R_b^2$ as in Eq.\ref{bar1}.
 Although this is an over simplification, the results will not be affected much as the second tunneling is
important only at deep sub-barrier energy where higher partial waves do not have significant contributions.

In the present model, the adiabatic neck barrier $V_n=V_m+V_{neck}(R_m,\nu_b)$ can be written as
\begin{eqnarray}
V_n^0= V_m+\frac{4\pi \gamma \bar R^2}{27} \sim V_m+0.46 \gamma \bar R^2.
\label{V_ad}
\end{eqnarray} 
As an example, for $^{64}Ni+^{64}Ni$ system, the neck barrier $V_{neck}(R_m,\nu_b)\sim 2.7$ MeV if we consider a nominal
$\gamma=1.0$ MeVfm$^{-2}$. Assuming $V_m \sim 85$ MeV for M3YR potential, the adiabatic barrier
will have a height of  about $V_n \sim 87.7$ MeV. Therefore, fusion cross section will be suppressed at a
threshold energy $E_0 \sim 87.7$ MeV which is consistent with the experimental observations. Similarly,
for light symmetric system like $^{16}O+^{16}O$, the neck potential $V_{neck}(R_m,\nu_b)$
 is about 1 MeV. This system should have
an adiabatic neck barrier of the order of 3.5 MeV where $V_m$ of 2.5 MeV has been assumed \cite{esben2}. 

Next we will examine whether fusion hindrance will occur for all projectile target combinations.
All along we have been arguing that the shape relaxation requires transition from $\nu=0$
to $\nu=\bar \nu$ which is separated by a barrier at $\nu_b$ (see Fig. \ref{fig4}).  
An important aspect which we have not considered yet is the magnitude of initial neck radius $\nu_g$
which is formed purely due to geometrical overlap of the diffused nuclear surfaces. In an approximate way
we can estimate an upper limit of $\nu_g$ by assuming that two times the  geometrical neck radius (the diameter 
of the cylinder) should not exceed the length of
the cylinder,
\begin{eqnarray}
 2 n_g \le l=(s+n_g^2/2 \bar R). 
\end{eqnarray}
This gives an acceptable solution,
\begin{eqnarray}
\nu_g=1-\sqrt{1-\rho}. 
\end{eqnarray}
for symmetric projectile target combinations. We generalise this expression 
for asymmetric combinations as well \cite{neck},
\begin{eqnarray}
\nu_g= (1-\sqrt{1-\rho})\frac{R_1+R_2}{4\bar R}.
\end{eqnarray}
 At $\rho=0.25$, $\nu_g$ turns out to be $\sim 0.13$ for all symmetric systems
which is less than
$\nu_b=1/6$. At $r=R_m$, the geometrical neck $\nu_g$ lies to the left of
$\nu_b$ and transition to $\bar \nu$ requires further tunneling. Therefore, for symmetric projectile
target systems, fusion hindrance seems to be a general phenomena at deep sub-barrier energies. However,
for asymmetric projectile target combinations, $\nu_g$ can be different depending on the asymmetry. If
$\nu_g < \nu_b$, further tunneling is required for shape relaxation which will lead to fusion hindrance.
On the otherhand, if $\nu_g > \nu_b$, there is no barrier to tunnel through and shape relaxation occurs
automatically. For example, in case of $^{6}Li+^{198}Pt$, $\nu_g \sim .19$ and lies to the right of
$\nu_b=1/6$. Thus, for this system, shape relaxation occurs automatically without any hindrance.
Using this model, we notice that $^{16}O+^{208}Pb$ system may undergo barrier tunneling in neck
direction.   
However, the present model is too simple to give any quantitative
prediction at which asymmetry supression will be absent.  This is only to 
demonstrate the mechanism for fusion hindrance which may not be found for highly
asymmetric combinations. The value of $V_{n}$ as given
in Eq.\ref{V_ad} is only an indicative and may not be valid for highly asymmetric system. Further,
the present  macroscopic calculation does not
include any shell correction which is very important at low excitation
energy. Therefore, in the actual coupled channel calculations, we will treat $V_{n}$
as a parameter which will be varied to fit the experimental measurements.

\section{Coupled Channel Calculation}
     
The above double barrier penetration model can be extended to calculate fusion cross section 
using coupled channel formalism. In this formalism,
the sudden tunneling probability $T^s$  is  estimated as a weighted sum over the tunneling
through a distributions of eigen barriers. This concepts was first introduced in \cite{brog} that
resulted in the computer code popularly known as CCFUS \cite{dasso}. We use the same CCFUS code with
suitable modification (CCFUSM) to incorporate the second tunneling phenomena. Since, we use the parabolic approximation,
the potential parameters of the model are $V_b$, $R_b$, $\hbar \omega$ (corresponding to the sudden barrier)
and $V_{n}$ and $\hbar \omega_{n}$ for adiabatic barrier. The sudden potential parameters are fixed
by fitting the fusion data around the Coulomb barrier where as adiabatic parameters are varied to fit
the data at deep sub-barrier energy.

\begin{figure}
\begin{center}
\includegraphics[scale=.4]{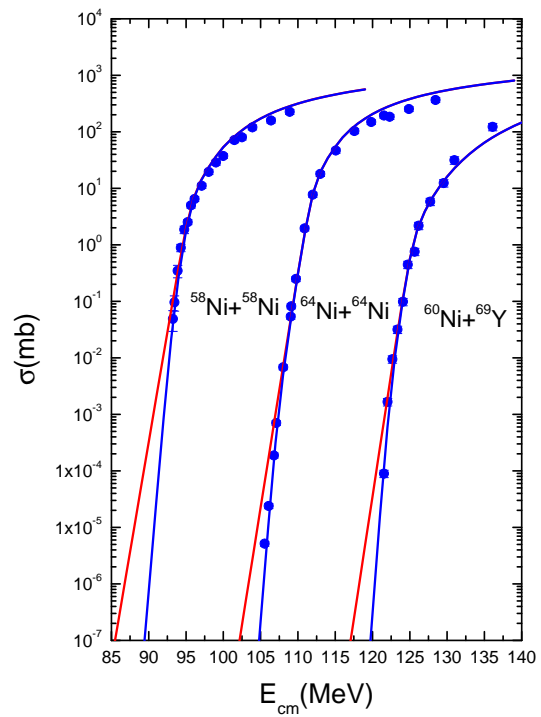}
\caption{ The fusion cross section $\sigma$ as a function  of $E_{cm}$ 
for $^{58}Ni+^{58}Ni$, $^{64}Ni+^{64}Ni$ and $^{60}Ni+^{69}Y$ systems.
The experimental data points are taken from \cite{beck,ack,jiang1,jiang2}. For $^{64}Ni+^{64}Ni$
system, the X-axis has been shifted by $+20$ MeV for clarity. The  red curves  are
obtained using standard coupled channel calculations. The coupling parameters are taken from
\cite{jiang6,brog}. The potential parameters are listed in table I.
 The blue curves  are the results of modified
coupled channel calculations with double tunneling factors. The corresponding adiabatic
parameters ($V_{n}$, $\hbar \omega_{n}$)are $93$ MeV and $3.5$ MeV for   
$^{58}Ni+^{58}$ system, $87$ MeV, $3.3$ MeV for $^{64}Ni+^{64}Ni$ system and $122$ MeV,
$3.1$ MeV for $^{60}Ni+^{69}Y$ system.} 
\label{fig5}
\end{center}
\end{figure}
\begin{figure}
\begin{center}
\includegraphics[scale=.4]{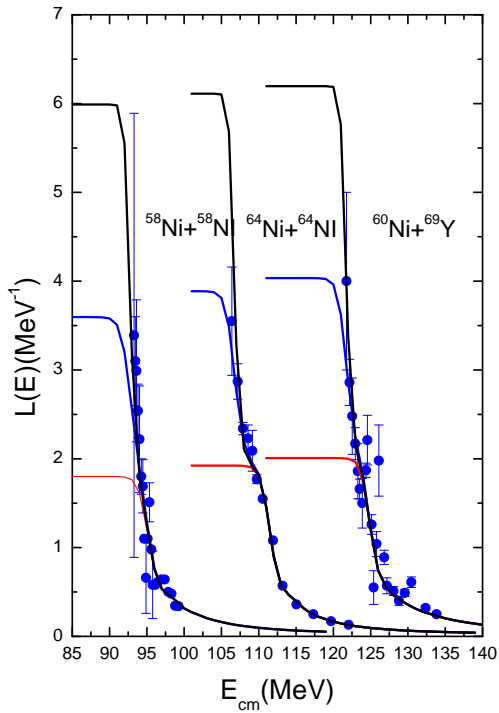}
\caption{ The corresponding logarithmic derivatives $L(E)$  as a function  of $E_{cm}$ 
for $^{58}Ni+^{58}Ni$, $^{64}Ni+^{64}Ni$ and $^{60}Ni+^{69}Y$ systems. All the curves have
same meaning as that of Fig. 5. The black curves
are obtained using a smaller $\hbar \omega_{n}$ of $1.5$ MeV.
}
\label{fig6}
\end{center}
\end{figure}
\begin{figure}
\begin{center}
\includegraphics[scale=.35]{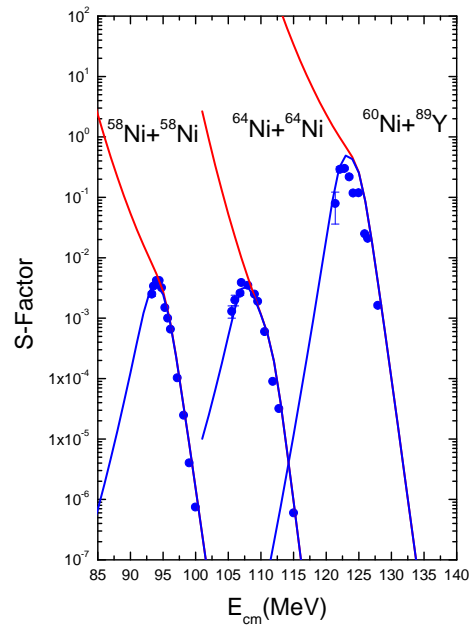}
\caption{ The corresponding $S$ factors  as a function  of $E_{cm}$ 
for $^{58}Ni+^{58}Ni$, $^{64}Ni+^{64}Ni$ and $^{60}Ni+^{69}Y$ systems. All the curves have 
same meaning as that of Fig. 5. The normalization factors $\eta_0$ are $69.99$, $75.23$ and $92.98$
respectively. 
}
\label{fig7}
\end{center}
\end{figure}

We consider three systems $^{58}Ni+^{58}Ni$, $^{64}Ni+^{64}Ni$ and $^{60}Ni+^{89}Y$ for which the anomalies
were reported for the first time \cite{jiang1,jiang2} although measurements have been carried out for
many other systems subsequently \cite{jiang3,jiang4,trotta}. These systems show a steep decrease of fusion
cross section at deep sub-barrier energy with increasing logarithmic slope $L(E)$ with decreasing energy.
Figs. (\ref{fig5},\ref{fig6}, \ref{fig7}) show the fusion 
cross section, logarithmic derivative and $S$ factors for the above
three systems. The red curves are the results obtained using the standard coupled channel calculations
(CCFUS) with potential parameters as listed in table I and also given in the figure captions. As can be seen, the coupled channel 
calculations reproduce the fusion cross sections near and above the Coulomb barrier quite well. However, it
overpredicts the measurements at deep sub-barrier energies. The $L(E)$ obtained from the coupled channel
calculations saturate to a nearly constant value at deep sub-barrier energies where as the experimental
$L(E)$ keeps on increasing with decreasing energies (See Fig. \ref{fig6}).

\begin{figure}
\begin{center}
\includegraphics[scale=.4]{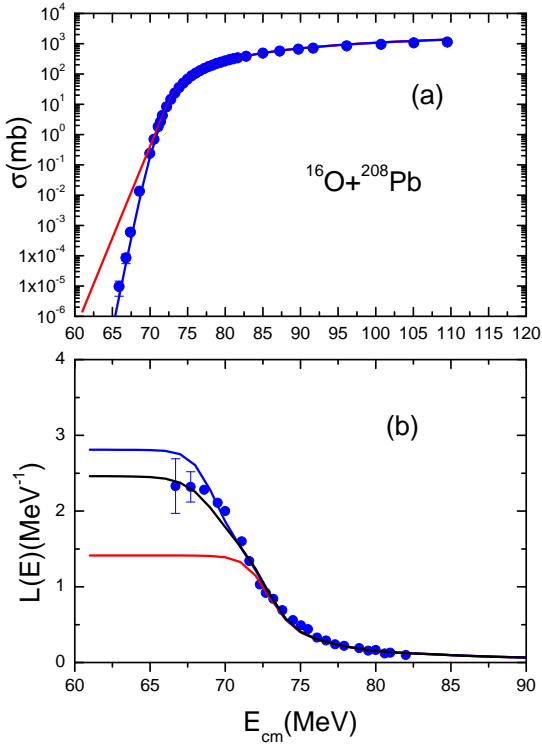}
\caption{ (a) The fusion cross section $\sigma$ as a function  of $E_{cm}$ 
for $^{16}O+^{208}Pb$ system.
The experimental data points are taken from \cite{das2,morton}.
The red curve is
obtained using standard coupled channel calculations. The coupling parameters are taken from
\cite{morton} and the potential parameters are listed in table I.
The blue curve  is the result of modified
coupled channel calculations with double tunneling factors. The adiabatic
parameters ($V_n$, $\hbar \omega_n$)are $69.5$ MeV and $4.5$ MeV respectively. (b)
The corresponding logarithmic derivatives $L(E)$  as a function  of $E_{cm}$ .
The black curve 
is obtained using a larger $\hbar \omega_{n}$ of $6.0$ MeV.
}
\label{fig8}
\end{center}
\end{figure}

\begin{figure}
\begin{center}
\includegraphics[scale=.4]{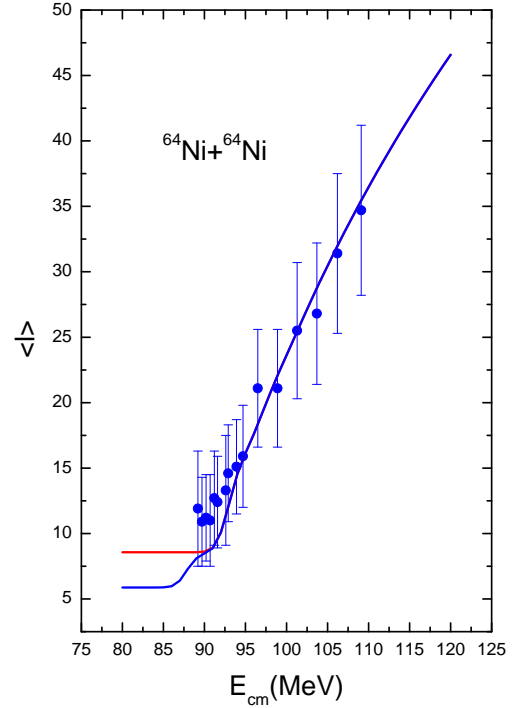}
\caption{ The average spin values $<l>$ as a function of $E_{cm}$ for
$^{64}Ni+^{64}Ni$ system. The red curve  is obtained using standard
coupled channel calculation where as the blue curve is obtained using modified
coupled channel calculation. The data points are taken from \cite{ack}.
}
\label{fig9}
\end{center}
\end{figure}

Similarly, as shown in Fig. \ref{fig7}, the $S$ factor keeps on increasing with decreasing energies where
as the experimental data show a well defined peak. This peak position correlates well  to the threshold
energy $E_0$ beyond which the experimental fusion cross section falls-off steeply. Next, we apply the
modified coupled channel (CCFUSM) calculation using a second tunneling factor as discussed previously. For
simplicity, we choose $\hbar \omega = \hbar \omega_{n}$, although these two parameters are completely
uncorrelated. The barrier height $V_n$ is adjusted to reproduce the experimental data at deep sub-barrier
energies. In fact, $V_n$ turns out to be same as that of $E_0$. The blue curves show the results of CCFUSM 
calculations which explain the data quite well. The $L(E)$ is quite sensitive to the $\hbar \omega_{n}$
parameter. As an example, we have used a smaller $\hbar \omega_{n}=1.5$ MeV for $L(E)$
calculations  for all the above three systems. The rise is still sharper as can be seen from the
black curves in Fig.\ref{fig6}. Therefore, $\hbar \omega_n$ is an important parameter and needs to be
estimated properly although we have set it to $\hbar \omega$ for simplicity.

We also consider $^{16}O+^{208}Pb$ system for which new measurements of fusion cross sections
are available at deep sub-barrier energies \cite{das2} complementing existing above barrier measurements
\cite{morton, hinde}. Unlike, $Ni$-induced reactions, the logarithmic derivative shows saturation
as shown in Fig. \ref{fig8}. The coupled channel calculations can not reproduce the data (see the red
curves). The blue curves show the results of modified coupled channel calculations with $V_{n}=69.5$ MeV
and $\hbar \omega=\hbar \omega_{n}=4.4$ MeV. The black curve in Fig.\ref{fig8}b is obtained using a higher
$\hbar \omega_n$ of $6$ MeV.
Note that the saturation phenomena can be understood using a suitable $\hbar 
\omega_{n}$ value. It may be mentioned here that in our model, $L(E)$ will ultimately lead to a saturation
value depending on the $\hbar \omega_{n}$ value. 

Another measurable quantity of interest is the average spin value which is defined as
\begin{eqnarray}
<l>=\frac{\sum l  \sigma_l}{\sigma_f}
\end{eqnarray}
Fig. \ref{fig9} shows $<l>$ as a function of energy for $^{64}Ni+^{64}Ni$ system. Although the results obtained
using normal coupled channel calculations and CCFUSM are identical at near and above barrier energies, the results
differ at deep sub-barrier energies.  

\section{Summary and Conclusion}   
We have proposed a hybrid coupled channel model for fusion using an ion-ion potential which has
same form as that of frozen density approximation at the nuclear exterior region and
becomes adiabatic in the nuclear interior region where the potential strongly depends on
both radial and neck degree of freedom. Ideally, the tunneling probability should have been
calculated through a two dimensional potential landscape. Instead, we follow a two step approach under the
assumption that initial radial motion is fast enough so that nuclear density is frozen at the
approach stage and after passage over the Coulomb barrier, the radial motion becomes
slower and the neck degree of freedom becomes important. Accordingly, the total tunneling probability is
written as the product of two factors, one for tunneling through the Coulomb barrier and the other one through
the barrier along the neck direction. For energy above and just below the Coulomb barrier, 
the tunneling probability along the neck barrier becomes almost unity and the hybrid model is just the normal
coupled channel model which is being used for fusion calculations. However, the second tunneling
probability becomes significant only at deep sub-barrier energy which becomes responsible for fusion
hindrance. It is also very important to use the right form of sudden ion-ion potential. The commonly used
AW parameterization does not include any short distance repulsive correction. On the otherhand, the M3YR
potentail is much thicker and shallower as the repulsive parameters have been adjusted to fit the deep
sub-barrier data under the assumption that shape of the ion-ion potential in the interior region decides
the fusion hindrance and neglects any other effects of dynamical origin like transition to di-nuclear to
mono-nuclear state through shape relaxation. Therefore, we believe, both AW and M3YR potentials are two extreme 
representations and the actual sudden ion-ion potential should follow the intermediate behavior. Therefore,
in this work we use a simple code like CCFUS which adopts parabolic approximation for tunneling calculations.
As a result, whatever be the form of potential, the tunneling probability is only sensitive to three parameters,
barrier height, barrier width and barrier position. Although, this is an over simplification particularly
at deep sub-barrier energy, we still follow the CCFUS approach to study the effect of
tunneling along the neck direction on fusion cross section without giving much emphasis on the shape
of the ion-ion potential in the nuclear interior region.  
Interestingly, it is found that the
magnitude of this adiabatic neck
barrier $V_{n}$ is quite close to the threshold energy $E_0$ and can
be written as $V_{n}=V_m+V_{neck}$ where $V_m$ is the minimum of the potential pocket of
the sudden potential and $V_{neck}$ is the additional barrier due to neck
formation which is estimated about $2$ MeV to $3$ MeV depending on the colliding systems. This finding
is consistent with the experimental observations. Our model also suggests that after a steep rise, the logarithmic
derivative will attain saturation  asymptotically. Another important finding of this model is that the
deep sub-barrier fusion hindrance is a general phenomena for all symmetric systems whereas hindrance
may not be seen for highly asymmetric  projectile target combinations, although it has not been possible
to give a asymmetric cut-off above which fusion hindrance will be found.

\begin{table}
\caption{The potential parameters for various systems used in the calculations.}
\begin{tabular}{|l|c|c|c|c|c|c|}\hline\hline
System  &~~~~$V_0$ & ~~~~$r_0$ &~~~~ $a_0$  &~~~~ $V_b$ &~~~~ $R_b$ &~~~~ 
$\hbar \omega$ \\
\hline
$^{64}Ni+^{64}Ni$  & 53.0 & 1.23 & 0.68 & 94.5 & 11.1 & 3.3 \\
\hline
$^{58}Ni+^{58}Ni$  & 53.0 & 1.204 & 0.68 & 99.6 & 10.5  & 3.5 \\
\hline
$^{60}Ni+^{89}Y$  & 53.0 & 1.178 & 0.68 & 134.8 & 10.7 & 3.1 \\
\hline
$^{16}O+^{208}Pb$  & 40.0 & 1.25 & 0.63 & 75.2 & 11.8 & 4.4 \\
\hline
\end{tabular}
\end{table}



\begin{thebibliography}{99}
\bibitem{neck} For asymmetric target projectile combinations of radius $R_1$ and $R_2$, we can write
               $n_g \sim (R_1 \nu_g + R_2 \nu_g)/2$. Dividing by $2 \bar R$, 
               we can have $\nu_g=\nu_g^s (R_1+R_2)
               /(4 \bar R)$, where $\nu_g^s=(1-\sqrt{1-\rho})$.
\end{thebibliography}
\end{document}